\documentclass[%
 reprint,
 amsmath,amssymb,
 aps,
 pre,
]{revtex4-2}
\usepackage{mathtools}
\usepackage{physics}
\usepackage{braket}
\usepackage{here}
\usepackage{graphicx}
\usepackage{dcolumn}
\usepackage{bm}
\usepackage{newtxtext,newtxmath}
\usepackage{hyperref}
\hypersetup{
setpagesize=false,
bookmarksnumbered=false,%
bookmarksopen=false,%
colorlinks=true,%
linkcolor=blue,
citecolor=blue,
urlcolor=blue,
}
\usepackage[normalem]{ulem}
\usepackage{xcolor}
\usepackage{lineno}
\usepackage[percent]{overpic}

\begin{document}

\title{Localized inhomogeneity and position-dependent stability of migratory bird formations}

\author{Hui Jiang}
\author{Nariya Uchida}%
\email{nariya.uchida@tohoku.ac.jp}
\affiliation{%
Department of Physics, Tohoku University, Sendai 980-8578, Japan
}%

\date{\today}

\begin{abstract}
We investigate how localized inhomogeneity affects the geometry and
stability of migratory bird formations. We use a lifting-line model
with a horseshoe-vortex representation to describe the longitudinal
dynamics of aerodynamic interactions. As a reference case, we first
analyze homogeneous formations and show that their steady states exhibit
a U-shaped geometry with hierarchical streamwise spacing, in which
adjacent birds become progressively closer toward the leader. We then
introduce localized inhomogeneity by modifying the wingspan of a single
bird, with its physical properties determined by scaling relations. We
determine the range of wingspan variation that preserves a stable
formation. The stability range depends strongly on the position of the
modified bird, being narrower near the outer wing and broader near the
leader. These findings provide a minimal dynamical framework for
understanding how local aerodynamic interactions and localized
individual differences affect collective flight structures.
\end{abstract}


\maketitle


\section{\label{sec:introduction}Introduction}

Migratory birds often travel in highly organized flight formations
\cite{Voelkl2017relation,andersson2004kin,bajec2009organized,ward1978formation,heppner1974avian}.
Such formations enhance flight efficiency by reducing energy expenditure
during long-distance migration
\cite{nathan2008vlike,sewatkar2010first,beauchamp2011long,lissaman1970formation,badgerow1981energy,weimerskirch2001energy}.
Various biological mechanisms have been proposed,
including visual coordination
\cite{heppner1985visual,li2017v-shaped,seiler2002analysis},
collective navigation
\cite{dorst1962migration,hamilton1967social},
and predation avoidance
\cite{vine1971risk}.
Among these, the aerodynamic mechanism provides a quantitative framework
in which energy savings arise from vortex-mediated interactions between individuals.
Birds exploit the upwash generated by these vortices
by positioning themselves within the wake of preceding birds,
thereby reducing induced drag
\cite{cattivelli2011modeling,lissaman1970formation,cutts1994energy,portugal2014upwash}.
This aerodynamic effect is widely regarded as a key factor
underlying the emergence and maintenance of formation flight
\cite{hummel1995formation,corcoran2019compound}.

Experimental observations support these aerodynamic advantages.
Measurements of heart rate \cite{weimerskirch2001energy}, wingbeat kinematics \cite{portugal2014upwash}, and relative positioning \cite{Hainsworth1987},
as well as three-dimensional tracking using GPS \cite{portugal2014upwash,voelkl2015matching,perinot2023characterization},
show that birds in formation reduce energetic costs
and occupy positions consistent with aerodynamic predictions.
Numerical studies, including computational fluid dynamics simulations,
reveal the vortex structures and wake interactions underlying these effects
\cite{billingsley2021role,maeng2013modeling,beaumont2023aerodynamic,beaumont2025aerodymic,beaumont2025aerodynamics}.
However, existing approaches either rely on rule-based
and control-oriented frameworks
\cite{vicsek1995novel,couzin2002collective,reynolds1987,cucker2007emergent,herbertRead2016understanding,cattivelli2009self}, or focus on
equilibrium configurations and their stability \cite{sugimoto2001theoretical}.
Consequently, the dynamical evolution of formation geometry
arising directly from aerodynamic interactions remains largely unexplored.

In reality, migratory flocks exhibit substantial variability
in body size and aerodynamic properties.
Despite this, heterogeneous formations have received limited attention.
Previous studies have considered such variability in prescribed configurations~\cite{thien2007effects,hummel1983aerodynamic}
or through rule-based adaptive models~\cite{cattivelli2011modeling}, and show that coherent
formations can still arise in heterogeneous groups.
More recent work has incorporated variability, typically through variations in wingspan,
and examined the resulting changes in formation geometry and energy saving~\cite{mirzaeinia2020flocking}.
Nevertheless, these studies do not capture how localized inhomogeneity
shapes formation geometry and dynamical stability
through aerodynamic interactions.

The geometry of formation flight is not restricted to a simple V-shaped configuration.
Using a nonlinear dynamical formulation based on lifting-line theory
with an elliptically distributed spanwise loading,
it was shown that self-organized steady formations
emerge as stable solutions, including U-shaped equilibrium configurations~\cite{sugimoto2001theoretical}.
From a different perspective, previous studies compared V and U formations
and showed that U-shaped configurations yield a more uniform distribution
of aerodynamic benefits and are energetically favorable~\cite{kawabe2007study}.
These studies characterize formation geometry
in terms of equilibrium states or optimal configurations,
but do not address the dynamical processes
through which such structures emerge and are stabilized.

We investigate the formation flight of migrating birds using a minimal dynamical model based on lifting-line theory with a single horseshoe-vortex representation, in which each wing is modeled as an effective bound vortex segment with trailing vortices.
The model is coupled to longitudinal flight dynamics and focuses on the predominantly local character of aerodynamic interactions through a simplified local-interaction approximation. For homogeneous formations, we obtain steady states with hierarchical spacing, in which adjacent birds become progressively closer toward the leader, yielding a U-shaped geometry with hierarchical streamwise spacing. Introducing inhomogeneity through a localized modification of wingspan, we demonstrate that the stability of the formation depends strongly on the position of the modified bird. In particular, the stability range is broader near the leader and narrower near the outer wing. These findings provide a minimal dynamical framework for understanding how local aerodynamic interactions and localized inhomogeneity affect the emergence and maintenance of collective flight structures.

\section{\label{sec:model}Model}

In the present study, the surrounding air is treated as incompressible and inviscid within the standard lifting-line framework, following previous theoretical studies of formation flight~\cite{sugimoto2001theoretical}.
Viscous effects are incorporated phenomenologically through the drag force.
We employ a quasi-steady aerodynamic description that captures the dominant aerodynamic interactions while neglecting unsteady wake dynamics.

\subsection{Solo flight}

We consider a migrating bird of mass $m$ flying 
at a speed $U(t)$ in the $x$ direction.
The equation of motion has the form
\begin{align}
m \frac{dU}{dt} = T - D 
\label{eq:eom1}
\end{align}
where $T$ is the thrust force 
and $D$ is the aerodynamic drag force.
The thrust force averaged over a flapping cycle
obeys the scaling relation~\cite{pennycuick2008modelling, shyy2010recent},
\begin{align}
T \approx C_T \rho S (2\pi f A)^2
\label{eq:thrust}
\end{align}
where $\rho$ is the air density, $S$ is the wing planform area, $f$ is the flapping frequency, $A$ is 
the wingtip amplitude, 
and $C_T$ ($\approx 0.1-0.4$) 
is the dimensionless thrust coefficient.
The drag force for a steady level flight is decomposed
into the profile drag and induced drag as
\begin{align} 
D &= k_{\rm pro} U^2 + k_{\rm ind} U^{-2},
\label{eq:D}
\\
k_{\rm pro} &= \frac12 C_{D} \rho S, 
\label{eq:kpro}
\\
k_{\rm ind} &= \frac{W^2}{2\pi e_{\rm O} \rho b^2}, 
\label{eq:kind}
\end{align}
where $C_{D}$ ($\approx 0.02-0.05$) 
is the zero-lift drag coefficient, 
$e_{\rm O}$ ($\approx 0.75-0.9$)
is the Oswald aerodynamic efficiency, 
and 
$W = mg$ is the weight of a bird of mass $m$
\cite{Tobalske2007, pennycuick2008modelling}.

The minimal-drag speed $U_{\rm min}$ is obtained 
by minimizing $D(U)$ as 
\begin{align} 
U_{\rm min}  
&= \left(\frac{k_{\rm ind}}{k_{\rm pro}}\right)^{1/4}
= 
\left(
\frac{{\rm AR}}{4 \pi e_{\rm O} C_D}
\right)^{1/4}
\left( \frac{W}{\rho b^2} \right)^{1/2},
\label{eq:Umin}
\\
D_{\rm min} &= D(U_{\rm min}) 
= 
2 \left( k_{\rm pro} k_{\rm ind} \right)^{1/2}
=
\left(
\frac{4 C_D}{\pi e_{\rm O} {\rm AR}}
\right)^{1/2} W,
\label{eq:Dmin}
\end{align}
where 
${\rm AR} = (2b)^2/S$ ($\approx 6-12$) 
is the aspect ratio.

Under isometric scaling, the geometric and physical quantities scale with the semi-wingspan $b$ as
$S \propto b^2$, $A\propto b$, $m \propto b^3$, 
and therefore
\begin{align} 
T \propto b^4, 
k_{\rm pro} \propto b^2, 
k_{\rm ind} \propto b^4,
U_{\rm min} \propto b^{1/2},
D_{\rm min} \propto b^{3}.
\label{eq:scaling}
\end{align}

The solo cruising speed $U_0$ of migrating birds is slightly 
(typically 5-15\%) larger than the minimum-drag speed.
We 
expand $D(U)$ around $U_0$ as
\begin{align} 
D(U) \simeq D_0 + D_1(U-U_0), 
\label{eq:DUlinear}
\end{align}
where $D_0 = D(U_0)$ and $D_1 = D'(U_0)$.
In a steady flight at $U=U_0$, 
the thrust force is balanced with the drag:
\begin{align} 
T = T_0 = D_0.
\label{eq:T0}
\end{align}

\subsection{Horseshoe vortex}

We consider a finite straight vortex segment ${\rm P}_1{\rm P}_2$ connecting 
the points ${\rm P}_1$:$(x_1, 0)$ and ${\rm P}_2$:$(x_2,0)$ ($x_1 > x_2$). 
According to the Biot–Savart law, the vertical velocity induced 
by the vortex segment at a point ${\rm Q}$:$(x, y)$ 
is given by
\begin{equation}
v_z(x,y) = -\frac{\Gamma}{4\pi y}
\left(\cos \varphi_1 + \cos \varphi_2 \right),
\label{eq:vz1}
\end{equation}
where $\Gamma$ is the circulation and 
$\varphi_i$ ($i=1,2$) is the angle between the vortex segment and ${\rm P}_i {\rm Q}$. 
We adopt the sign convention that $v_z>0$ corresponds to upwash 
and $v_z<0$ corresponds to downwash.

The horseshoe-vortex model represents each wing by a single effective bound vortex segment and two semi-infinite trailing vortices, replacing the full spanwise loading distribution with a reduced description of the finite-wing lift.
We assume that the wing is centered at the origin and that
the bound vortex segment connects ${\rm A}_+$:$(0,a)$ and ${\rm A}_-$:$(0,-a)$,
where $a$ is related to the semi-wingspan $b$ as
\begin{equation}
a = \frac{\pi}{4} b.
\end{equation}
We assume that the bird flies in the positive 
$x$-direction, 
and the trailing vortices extend from ${\rm A}_{\pm}$ toward the negative $x$ direction.
The circulation $\Gamma$ is related to the lift force $L$ via 
the Kutta–Joukowski theorem as $L = 2a \rho \Gamma U$.
In steady level flight, we set $L=W$. Using Eq.~(\ref{eq:Umin}), we obtain
\begin{align}
\Gamma &= \frac{W}{2a \rho U} 
= \frac{\Gamma_{\rm min} U_{\rm min}}{U},
\label{eq:Gamma}
\\ 
\Gamma_{\rm min} 
&=  
4
\left(\frac{e_{\rm O} C_D}{\pi {\rm AR}} 
\right)^{1/2} b U_{\rm min}.
\label{eq:Gammamin}
\end{align}  

The vertical velocity induced by the horseshoe vortex at the point ${\rm Q}$:$(x,y)$
is the sum of the contributions from the bound and trailing vortices.
For convenience, we define the vertical velocity $v_z$ divided 
by $-\Gamma/(4\pi)$ as $\hat{v}_z$.
The bound vortex gives 
\begin{align}
\hat{v}_{z,0} (x,y) &=  
\frac{1}{x}
\left[
\frac{y+a}{\sqrt{x^2+(y+a)^2}} - 
\frac{y-a}{\sqrt{x^2+(y-a)^2}} 
\right],
\label{eq:vz0}
\end{align}
and the trailing vortices give
\begin{align}
\hat{v}_{z, \pm} (x,y) &=
\frac{1}{a \pm y}
\left[
1 + \frac{x}{\sqrt{x^2+(y\pm a)^2}}  
\right].
\label{eq:vzpm}
\end{align}

\subsection{Aerodynamic interaction}

We consider a second bird centered at ${\rm Q}$:$(x,y)$ and moving in parallel with the first bird, which has a different size and circulation $\Gamma'$.
The bound vortices of the second bird are thus located at ${\rm B}_{\pm} $:$(x, y\pm a')$.
The signed interaction contribution to the induced drag experienced by
the second bird due to the first is given by
\begin{align}
D_{12} &= \frac{\rho \Gamma \Gamma'}{4\pi} I(x,y, a,a'),
\label{eq:D12}
\end{align}
where \(I(x,y,a,a')\) is obtained
by integrating the dimensionless induced velocity
\(\hat v_z\) generated by the first bird
over the wing span of the second bird \({\rm B_-B_+}\).
We obtain
\begin{align}
I &= I_0 + I_+ + I_-,
\label{eq:I}
\\
I_0 &= \int_{-a'}^{a'} ds \, \hat{v}_{z,0}(x, y+s)
\nonumber\\
&=  
\frac{1}{x}
\bigg[ J(x, y+a'+a) + J(x, y-a'-a) 
\nonumber\\
& \hspace{5mm} - J(x, y+a'-a) - J(x, y-a'+a) \bigg],
\label{eq:I0}
\\
I_\pm &=  \pm \int_{-a'}^{a'} ds \, \hat{v}_{z,\pm}(x, y+s)
\nonumber\\
&=  \pm
 \bigg[ \ln\left|\frac{y+a'\pm a}{y-a'\pm a}\right|
+K(x,y+a'\pm a)-K(x,y-a'\pm a) \bigg].
\label{eq:Ipm}
\end{align}
where we used the functions
\begin{align}
J(x,y) &= \sqrt{x^2 + y^2},
\\
K(x,y) &= 
\frac{x}{2|x|}
\ln\left|
\frac{\sqrt{x^2+y^2}-|x|}
{\sqrt{x^2+y^2}+|x|}
\right|.
\end{align}

\subsection{Formation flight}

\begin{figure}[t]
\centering
\includegraphics[width=0.8\columnwidth]{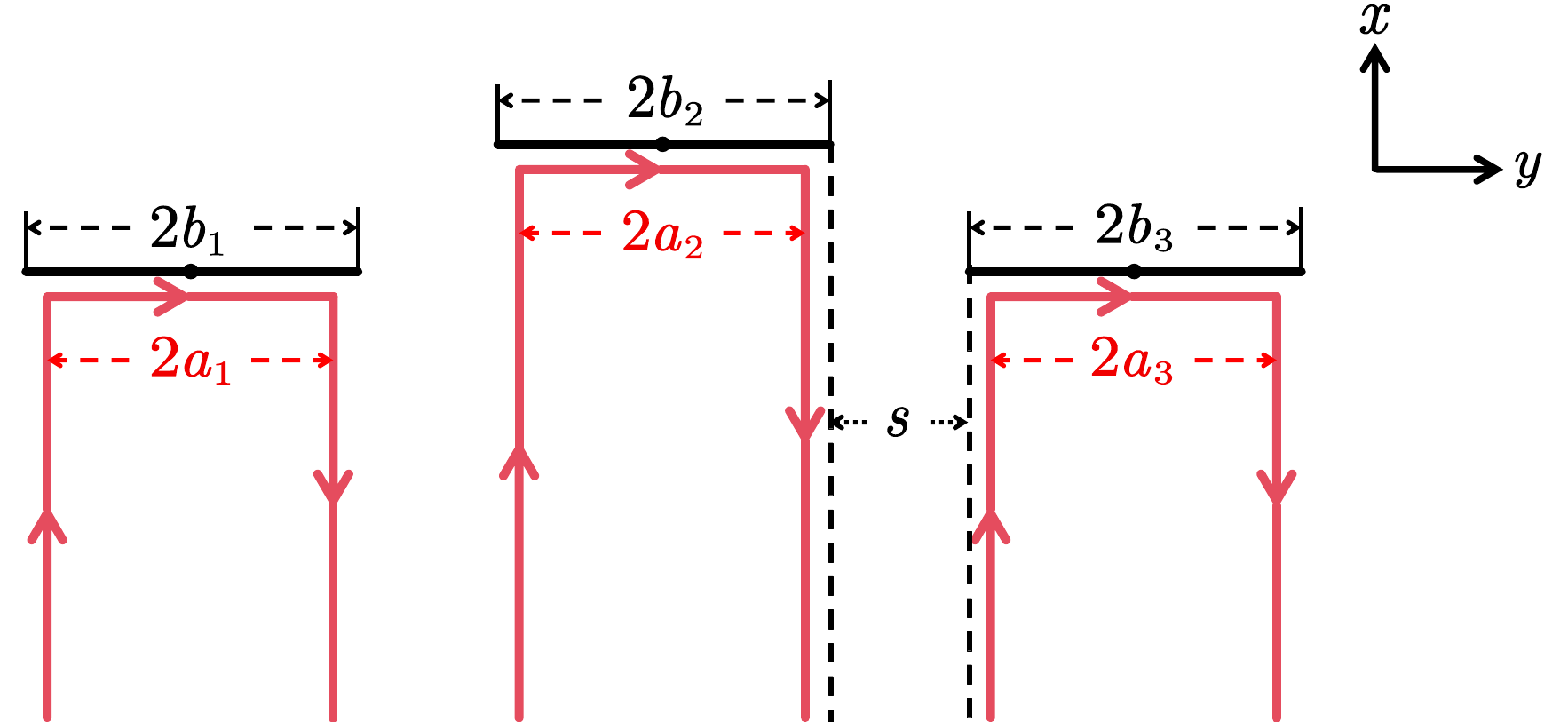}
\caption{
Local geometric definition of adjacent birds in the formation model.
For bird $i$, the full wingspan is $2b_i$, while the effective bound-vortex length is $2a_i$.
The streamwise and transverse separations between neighboring birds are denoted by $\Delta x$ and $\Delta y$, respectively.
The coordinate system is defined such that 
the positive $x$ direction is the flight direction
and $y$ is the transverse direction.
}
\label{fig:1}
\end{figure}

Previous studies~\cite{badgerow1981energy} have shown that optimal aerodynamic benefit occurs when the wingtip of a bird laterally overlaps with that of the bird in front.
They suggested an optimal wingtip spacing
$s_{\rm opt}=\left(\frac{\pi}{4}-1\right)b$.
Accordingly, for an adjacent pair of birds with semi-wingspans
$b_i$ and $b_j$, we set the transverse center-to-center separation to
\begin{equation}
\Delta y_{\rm adj} = b_i+b_j+s_{\rm opt}.
\end{equation}

We now consider the collective dynamics of a formation consisting of $N$ birds,
all moving in the positive $x$ direction.
We denote the center position of the $i$-th wing as $(x_i,y_i)$,
with $y_1<y_2<\cdots<y_N$.
The bird $i$ has circulation $\Gamma_i$, and the separation between its trailing vortices is $2a_i$.
Following Eq.~(\ref{eq:D12}),
the signed interaction-induced drag acting on bird $j$
due to bird $i$ is written as
\begin{align}
D_{ij} &= \frac{\rho \Gamma_i \Gamma_j}{4\pi} 
I(x_{ij}, y_{ij}, a_i, a_j), 
\label{eq:Dij}
\end{align}
where $x_{ij} = x_i - x_j$ and $y_{ij} = y_i - y_j$.
With this convention, \(D_{ij}<0\) corresponds to a drag-reducing interaction.

The equation of motion for the bird $j$ reads
\begin{align}
\frac{dx_j}{dt} &= U_j,
\\
m_j \frac{dU_j}{dt} &= T_j - D_j - \sum_{i\neq j} D_{ij},
\label{eq:eom2}
\end{align}
where $m_j$, $U_j$, $T_j$, and $D_j$ are the mass, 
velocity, thrust force and self-induced drag of the bird $j$, respectively.

\subsection{Scaling and inhomogeneity}

First, we rescale all the quantities by
the length $b$, time $\tau=b/U_{\rm min}$, 
and mass $m_{\rm air} = \rho b^3$,
assuming that all the birds have the same size 
and minimal-drag speed.
It gives the unit of force 
\begin{align}
F_{\rm min} 
= \rho b^2 U_{\rm min}^2 
= \frac{{\rm AR}}{4C_D} D_{\rm min}.
\label{eq:Fmin}
\end{align}
The dimensionless equation of motion reads 
\begin{align}
\frac{d\hat{x}_j}{d\hat{t}} &= \hat{U}_j,
\\
\hat{m}_j \frac{d\hat{U}_j}{d\hat{t}} &= \hat{T}_j - \hat{D}_j
- \sum_{i \neq j} \hat{D}_{ij},
\label{eq:eom3}
\end{align}
where $\hat{x}_j = x_j/b$, 
$\hat{U}_j = U_j/U_{\rm min}$,
$\hat{m}_j = m_j/m_{\rm air}$,
$\hat{T}_j = T_j/F_{\rm min}$, 
$\hat{D}_j = D_j/F_{\rm min}$, and
$\hat{D}_{ij} = D_{ij}/F_{\rm min}$.
The solo drag is expressed as
\begin{align}
\hat{D}_j &= 
\hat{k} 
\left( \hat{U}_j^2 + \hat{U}_j^{-2} \right),
\label{eq:hatD}
\\
\hat{k}
&= \frac{k_{\rm pro}}{\rho b^2}
= \frac{k_{\rm ind}}{\rho b^2 U_{\rm min}^4}
= \frac{2 C_D}{{\rm AR}}
\label{eq:hatk}
\end{align}
The interaction term 
becomes 
\begin{align}
\hat{D}_{ij} &= 
\frac{\hat{\Gamma}_i \hat{\Gamma}_j}{4\pi} 
I(\hat{x}_{ij}, \hat{y}_{ij}, \hat{a}_i, \hat{a}_j), 
\label{eq:Djk2}
\end{align}
where the dimensionless circulation is obtained
from Eqs.~(\ref{eq:Gamma}) and~(\ref{eq:Gammamin})
as
\begin{align}
\hat{\Gamma}_j 
= \frac{\Gamma_j}{bU_{\rm min}} .
\label{eq:hatGamma}
\end{align}

Next, we consider an inhomogeneous flock
with the semi-wingspan $b_j = \lambda_j b$   
for the bird $j$, where $b$ is the standard size
of the bird which has the mass 
$m_0 = \rho_{\rm bird} b^3$ 
and the minimal drag speed and force given by 
Eqs.(\ref{eq:Umin}),(\ref{eq:Dmin}).
From the scaling relations (\ref{eq:scaling}),
we assume
\begin{align}
\hat{m}_j &= \lambda_j^3 \hat{m}_0,
\\
\hat{T}_j &= \lambda_j^4 \hat{T}_0, 
\\
\hat{\Gamma}_j &= \lambda_j^2 \frac{\hat{\Gamma}_{\rm min}}{\hat{U}_j},
\quad 
\hat{\Gamma}_{\rm min} = 
4
\left(\frac{e_{\rm O} C_D}{\pi {\rm AR}} 
\right)^{1/2} .
\\
\hat{D}_j &= \hat{k} \left( 
\lambda_j^2 \, \hat{U}_j^2 
+
\lambda_j^4 \, \hat{U}_j^{-2} 
\right).
\end{align}
The minimum-drag speed of bird $j$
is obtained 
as
\begin{align}
\hat{U}_{{\rm min},j} &=  \lambda_j^{1/2},
\end{align}
while
its solo cruising speed 
is obtained from the force balance equation
$\hat{T}_j = \hat{D}_j$ as
\begin{align}
\hat{U}_{0,j}
=
\left(
\lambda_j^2 H_0
+
\sqrt{\lambda_j^4 H_0^2-\lambda_j^2}
\right)^{1/2},
\quad
H_0 = \frac{ 
\hat{U}_0^2 + \hat{U}_0^{-2} 
}{2}.
\end{align}
Here we have chosen the higher-speed branch. A real solution on this
branch exists for \(\lambda_j H_0 \ge 1\). 
At the limiting value,
\(\hat{U}_{0,j}=\hat{U}_{{\rm min},j}=\lambda_j^{1/2}\), whereas for
\(\lambda_j H_0>1\) one has
\(\hat{U}_{0,j}>\hat{U}_{{\rm min},j}\).
In this higher-speed regime,
\(d\hat{D}_j/d\hat{U}_j>0\), and hence the corresponding
solo-flight equilibrium is linearly stable.

In the following sections, we work with the dimensionless variables
introduced in this subsection, and omit the hat notation for simplicity.

\section{Analytical Results for Homogeneous Formations}

\subsection{\label{sec:ana}Nearest-neighbor approximation and hierarchical spacing}

In general, the aerodynamic interaction includes contributions
from all other birds in the formation.
However, owing to the spatial decay of the induced velocity,
the interaction is expected to be dominated by nearby individuals.

\begin{figure}[h]
\includegraphics[width=0.8\columnwidth]{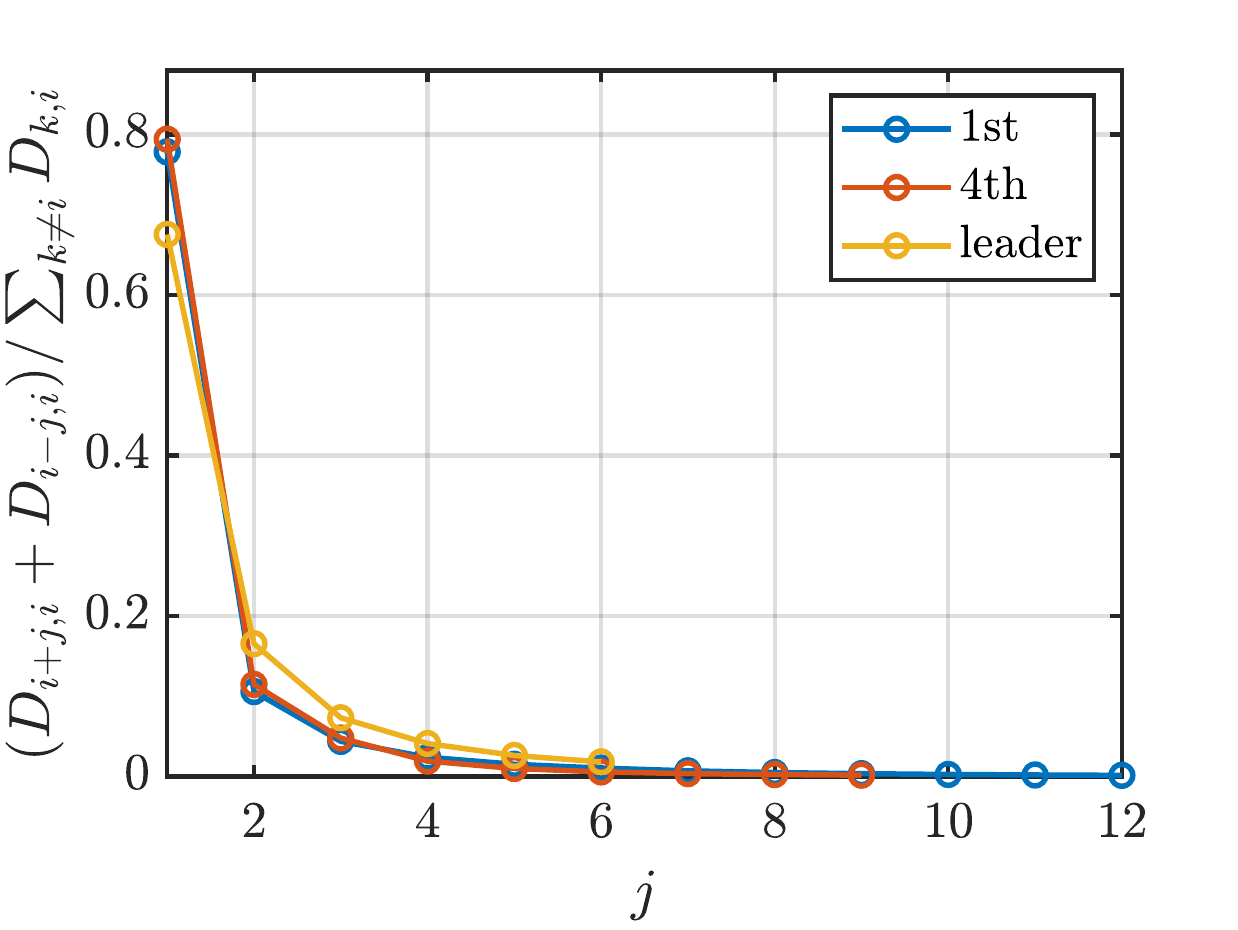}
\caption{
The normalized contribution
$(D_{i+j,i}+D_{i-j,i})/\sum_{k\ne i}D_{k,i}$
is shown as a function of the neighbor order $j$
for representative birds ($i=1,4,7$), corresponding to the outermost bird,
an intermediate bird on the left wing, and the leader, respectively.
\label{fig:2}
}
\end{figure}

To quantify this effect, we evaluate the contribution of different
neighbor orders to the total interaction acting on each bird,
using the pairwise interaction defined in Eq.~(\ref{eq:Dij}).
Figure~\ref{fig:2} shows the normalized contribution
$(D_{i+j,i} + D_{i-j,i}) / \sum_{k\ne i} D_{k,i}$
as a function of the neighbor order $j$.
The results indicate that the nearest-neighbor contribution
is significantly larger than higher-order contributions,
which decay rapidly with increasing separation.

This behavior indicates that the aerodynamic coupling is effectively
short-ranged.
Motivated by this observation, we employ a nearest-neighbor approximation
to obtain analytical insight into the formation structure.

We consider a symmetric formation consisting of an odd number of birds,
\begin{equation}
N=2n-1,
\qquad
n=\frac{N+1}{2},
\end{equation}
where $n$ denotes the number of birds on one side of the formation, including the leader.
Along one wing (from tail to leader), the birds are labeled by $i=1,2,\ldots,n$,
consistent with the coordinate system and local geometric definition
introduced in Fig.~\ref{fig:1},
and their streamwise positions satisfy
\begin{equation}
x_1 < x_2 < \cdots < x_n,
\label{eq:x_order}
\end{equation}
so that \(x_{i+1}-x_i>0\) gives the longitudinal spacing between adjacent birds.

For each adjacent pair along one wing, the sum of the two reciprocal
interactions,
$D_{i,i-1}+D_{i-1,i}$,
is independent of the streamwise separation.
In a steady homogeneous formation, all birds fly at a common speed $U^\ast$
and share a common circulation $\Gamma^\ast$.
Substituting $a'=a$ and $\Gamma_i=\Gamma_{i-1}=\Gamma^\ast$
into Eq.~(\ref{eq:Dij}),
the interaction associated with an adjacent pair reduces to
\begin{equation}
D_{i,i-1}+D_{i-1,i}
=
\frac{(\Gamma^\ast)^2}{2\pi}
\ln\!\left[
1-\left(\frac{2a}{\Delta y}\right)^2
\right]
\equiv C,
\qquad i=2,\ldots,n,
\label{eq:pairingC_x}
\end{equation}
where $C<0$ under the present sign convention, indicating a net drag-reducing interaction.
Here the transverse separation is fixed as
\begin{equation}
\Delta y = 2b + s_{\mathrm{opt}},
\end{equation}
where \(s_{\mathrm{opt}}=(\pi/4-1)b\)
is the optimal wingtip spacing~\cite{badgerow1981energy}.
The negative value of \(s_{\mathrm{opt}}\) corresponds to a wingtip overlap
of \(|s_{\mathrm{opt}}|\).

For a steady formation with constant inter-bird separations,
all birds fly at a common speed $U^\ast$.
In the homogeneous case, this implies identical thrust and self-induced drag,
$T_j=T^\ast$ and $D_j=D^\ast$ for all $j$.
The steady-state condition $dU_j/dt=0$ in Eq.~(\ref{eq:eom3}) then yields
\begin{equation}
\sum_{i\neq j} D_{ij} = T^\ast - D^\ast \equiv S,
\label{eq:S_def}
\end{equation}
where $S$ is independent of $j$.

Under the nearest-neighbor approximation, the force balance for internal birds becomes
\begin{equation}
D_{i-1,i}+D_{i+1,i}= S,
\qquad i=2,\ldots,n-1.
\label{eq:balance_internal}
\end{equation}
At the leader, the two neighboring birds contribute symmetrically, giving
\begin{equation}
2D_{n-1,n}=S.
\end{equation}

Combining Eqs.~(\ref{eq:pairingC_x})--(\ref{eq:balance_internal}),
we obtain a linear profile for the adjacent interactions along one wing,
\begin{equation}
D_{i,i-1}
=
C\left(1-\frac{i-1}{2n-1}\right),
\qquad i=2,3,\ldots,n.
\label{eq:linear_general_x}
\end{equation}
Since $C<0$ and the interaction decreases monotonically
with the longitudinal separation,
the magnitude $|D|$ increases monotonically with the inter-bird distance.
Equation~(\ref{eq:linear_general_x}) therefore implies
\begin{equation}
x_2-x_1 > x_3-x_2 > \cdots > x_n-x_{n-1},
\label{eq:ordering_general_x}
\end{equation}
showing that the longitudinal spacing between adjacent birds
is largest near the tail and decreases monotonically toward the leader.
\subsection{\label{sec:LSA} Linear stability analysis}

We analyze the linear stability of steady formations
in the homogeneous case by introducing small perturbations
to the equations of motion derived in Sec.~\ref{sec:model}.
Since the analytical treatment becomes increasingly complicated
as the number of birds increases, we focus here on the analytically
tractable case of three birds.

Owing to the symmetry of the three-bird steady formation,
the linear stability analysis can be reduced to two representative birds,
since the third bird obeys identical dynamics.
We introduce small perturbations
$\delta(x_2-x_1)$, $\delta U_1$, and $\delta U_2$
around the steady state, where the longitudinal separation is defined as
$x=x_2-x_1>0$ along the flight direction.

By performing first-order Taylor expansions of the interaction terms
$D_{12}$ and $D_{21}$ around the steady state,
in the same manner as in Eq.~(\ref{eq:DUlinear}),
we obtain the linearized equations of motion
\begin{align}
\frac{d\, \delta x}{d t} &= \delta U_2 - \delta U_1, \\[1ex]
\frac{d\, \delta U_1}{d t} &=
\frac{1}{m}
\left(
- Q\, \delta U_1
-
P\, \delta x
\right), \\[1ex]
\frac{d\, \delta U_2}{d t} &=
\frac{1}{m}
\left(
- Q\, \delta U_2
+
2P\, \delta x
\right).
\end{align}

Here $P$ characterizes the linear response of the pairwise aerodynamic interaction
to a perturbation in the longitudinal separation and is defined as
\begin{equation}
P \equiv
\left.
\frac{\partial D_{21}}
{\partial x}
\right|_{x=x^\ast},
\qquad
\left.
\frac{\partial D_{12}}
{\partial x}
\right|_{x=x^\ast}
= -P,
\end{equation}
where $x^\ast$ is the steady-state longitudinal separation.

The coefficient $Q$ is the drag restoring coefficient associated with
the self-induced drag, defined as
\begin{equation}
Q \equiv
\left.
\frac{\partial D_1}{\partial U_1}
\right|_{U_1=U^\ast}
=
\left.
\frac{\partial D_2}{\partial U_2}
\right|_{U_2=U^\ast}.
\end{equation}

Here $U^\ast$ is the common steady speed of the formation.
Since the steady speed $U^\ast$ is larger than the minimum-drag speed,
the slope of the drag curve is positive, yielding $Q>0$.

The linearized equations can be written in matrix form as
\begin{equation}
\frac{d}{d t}
\begin{bmatrix}
\delta x \\
\delta U_1 \\
\delta U_2
\end{bmatrix}
=
\mathbf{A}
\begin{bmatrix}
\delta x \\
\delta U_1 \\
\delta U_2
\end{bmatrix}.
\end{equation}

The eigenvalues of $\mathbf{A}$ are
\begin{equation}
\mu_1 = -\frac{Q}{m}, \quad
\mu_{2,3} = \frac{1}{2m}
\left(
- Q \pm \sqrt{Q^2 - 12mP}
\right).
\end{equation}
Since $Q>0$ and $m>0$, we have $\mu_1<0$.
For $\mu_{2,3}$, the real part is given by $-Q/(2m)<0$,
independently of the sign of $Q^2-12mP$.
Hence all eigenvalues have negative real parts,
and the steady three-bird formation is linearly stable.

For larger formations or inhomogeneous cases,
the linear stability analysis becomes analytically intractable.
We therefore assess the stability numerically in Sec.~\ref{sec:results},
where we introduce a quantitative stability criterion
and present the stability diagram for the $N=13$ case (Fig.~\ref{fig:4}).

\section{\label{sec:results} Numerical Results for Homogeneous Formations}

For the numerical simulations, the dimensionless parameters were chosen
based on representative physical values for Canada geese reported in
Ref.~\cite{mirzaeinia2020analytical}. 
We used the body mass \(M=3.8\,{\rm kg}\), full wingspan
\(B=1.50\,{\rm m}\), and minimum drag speed
\(U_{\rm min}=18.0\,{\rm m/s}\). 
The dimensionless mass of the
standard bird is
\[
m_0=\frac{M}{\rho_{\rm air}(B/2)^3} \approx 7.35,
\]
where \(\rho_{\rm air}=1.225\,{\rm kg/m^3}\).
The solo cruising speed was set to
\[
U_0=\frac{18.5}{18.0} \approx 1.028.
\]
In the nondimensional variables, the standard semi-wingspan is \(b=1\),
with \(a=\pi/4\) as defined in Sec.~\ref{sec:model}.
We use the wingtip spacing \(s_{\rm opt}=a-b=\pi/4-1\), as introduced
in Sec.~\ref{sec:ana}.

For the homogeneous simulations, the initial condition is a symmetric
V-shaped configuration with the leader placed at the origin. 
For \(N=2n-1\), the leader corresponds to \(i=n\), and the initial positions are
\[
x_i(0)=-|i-n|\Delta x_0,\qquad
y_i(0)=(i-n)\Delta y_0,
\]
where
\[
\Delta y_0=2b+s_{\rm opt},\qquad
\Delta x_0=\frac{\Delta y_0}{\tan\theta},
\qquad
\theta=\frac{\pi}{4}.
\]
All birds are assigned the same initial velocity,
\[
U_i(0)=U_0.
\]

To obtain steady formations, we numerically integrate the coupled
longitudinal equations of motion for all birds using a fourth-order
Runge--Kutta method with a fixed time step \(\Delta t=0.01\).
At each time step, the aerodynamic interaction acting on each bird is
evaluated from the instantaneous configuration based on the
horseshoe-vortex representation derived from the Biot--Savart law
(Sec.~\ref{sec:model}).

For homogeneous formations, we performed simulations both with the
all-to-all interaction and with the nearest-neighbor approximation in
order to assess the validity of the local-interaction approximation.
In the all-to-all calculations, the interaction terms were evaluated using the full pairwise geometric separations between all birds.
Unless otherwise stated, the following numerical
simulations use the nearest-neighbor approximation, reflecting the
short-range character of the aerodynamic interaction.

The system is evolved up to the final simulation time \(t_{\max}=10^5\),
by which the velocities and relative positions have converged to steady
values. The steady-state configuration and the corresponding flight speed
are determined from the long-time asymptotic state of the system.

\begin{figure}[h]
\begin{overpic}[width=0.90\columnwidth]{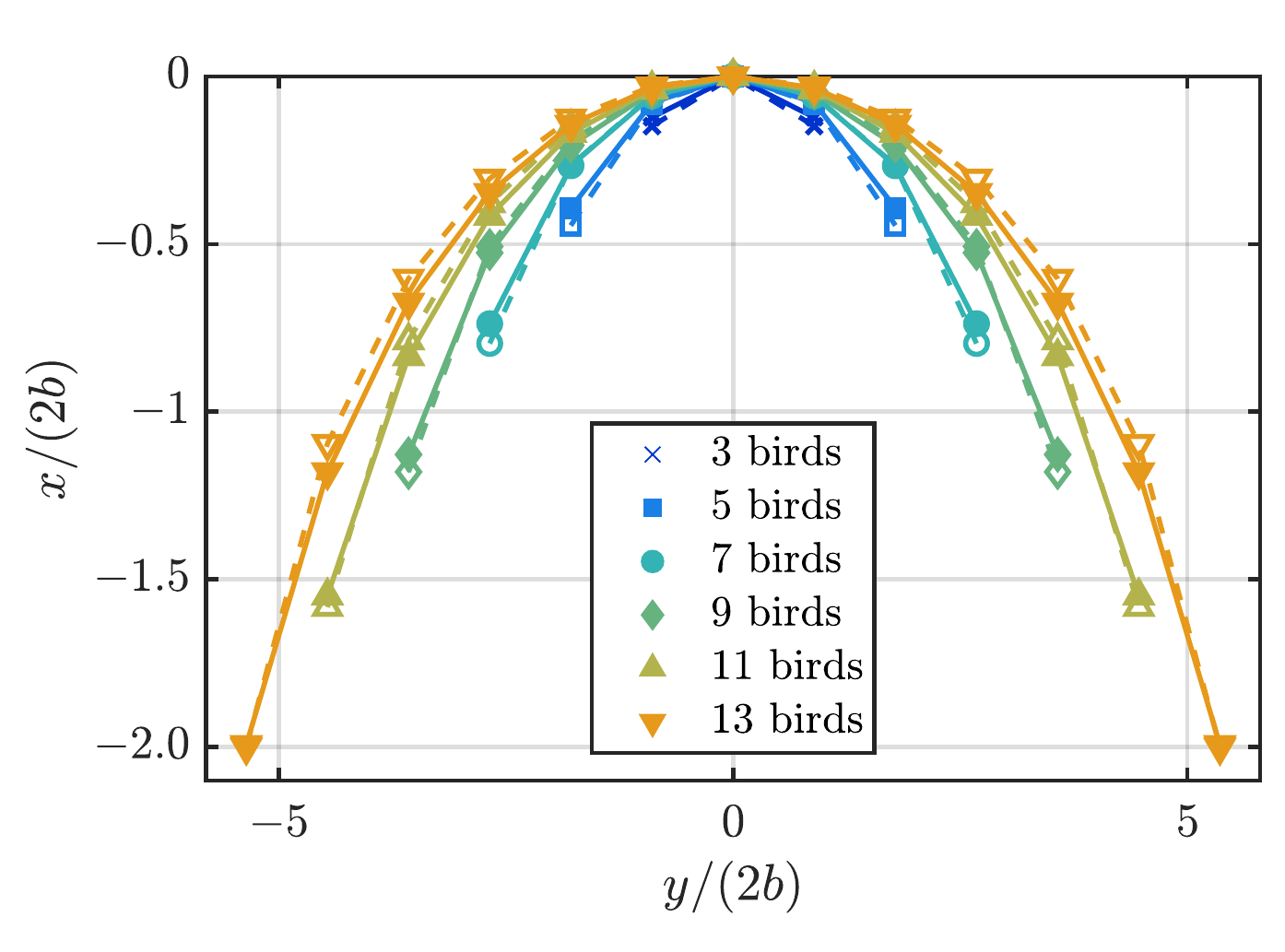}
\put(0,67){\large (a)}
\end{overpic}
\begin{overpic}[width=0.90\columnwidth]{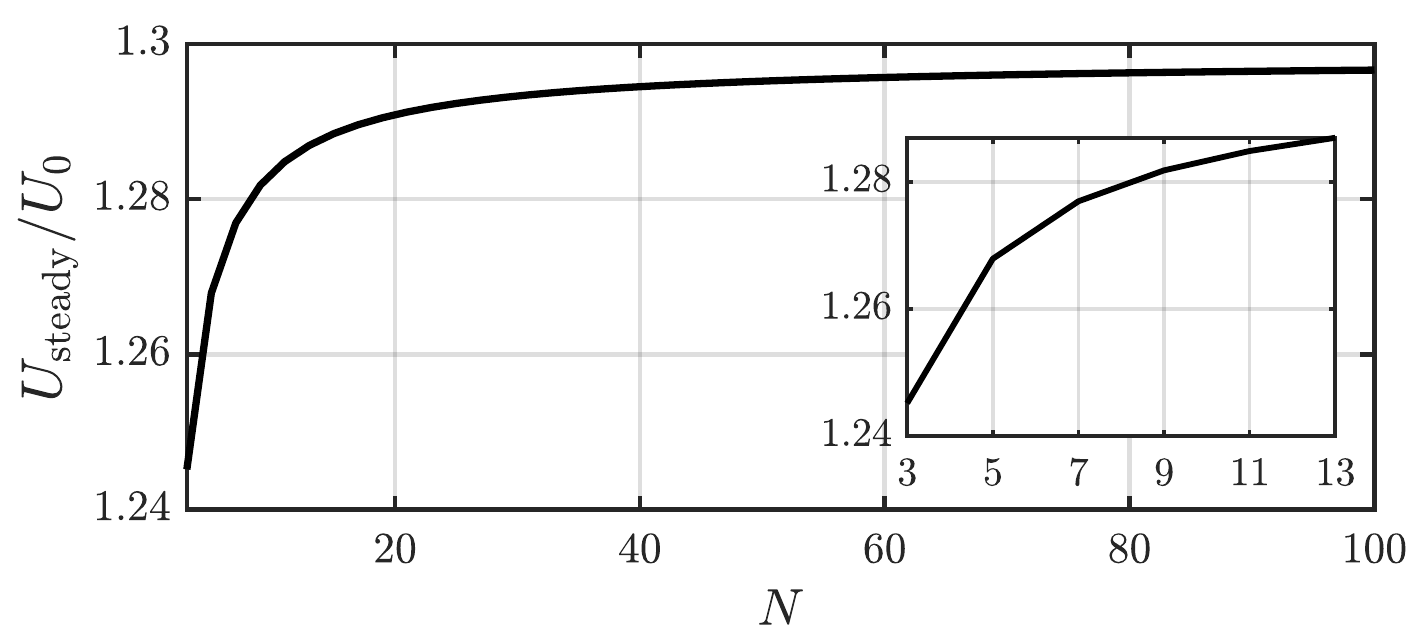}
\put(0,40){\large (b)}
\end{overpic}
\begin{overpic}[width=0.90\columnwidth]{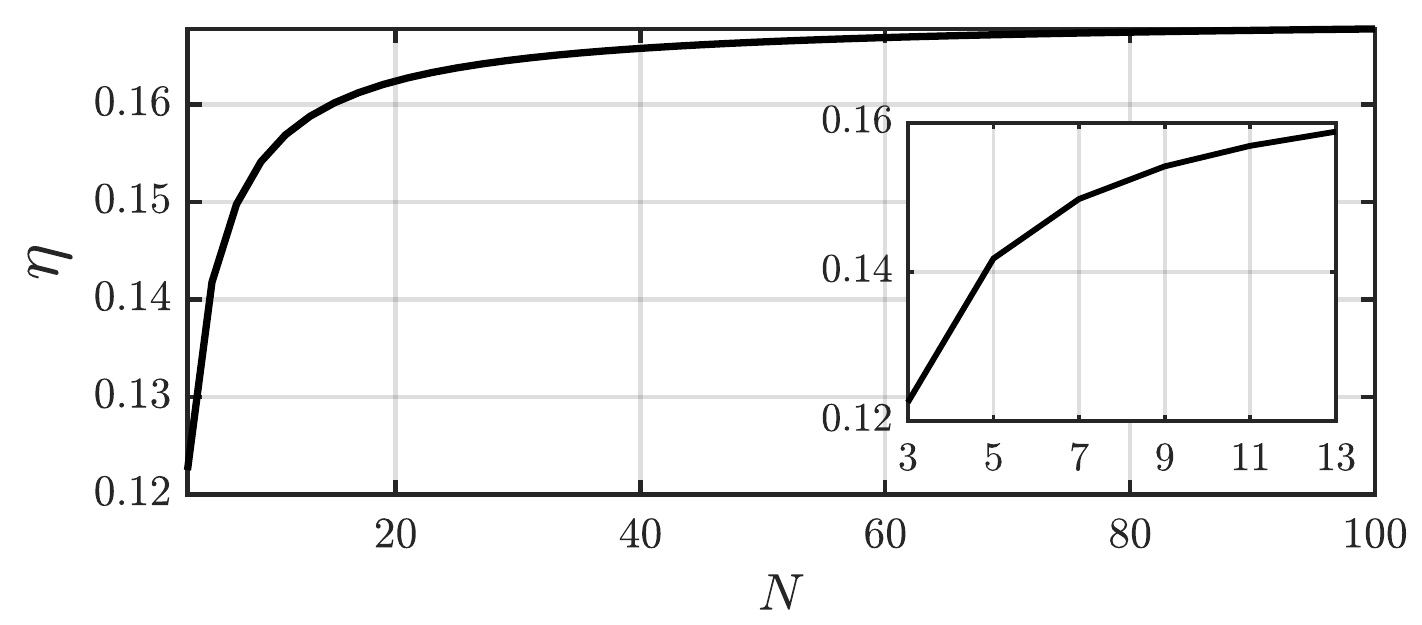}
\put(0,41){\large (c)}
\end{overpic}
\caption{
(a)
Steady-state configurations for homogeneous formations with different numbers of birds. 
The coordinates are scaled by the wingspan $2b$
and shifted so that the leader is located at the origin. 
Different symbols denote different flock sizes, while solid and dashed lines represent the all-to-all interaction model and the nearest-neighbor approximation, respectively.
(b)
Dependence of the steady-state velocity on the number of birds.
The inset highlights the cases corresponding to the configurations shown in Fig.~\ref{fig:3}(a).
(c)
Dependence of the drag reduction rate $\eta$ on the number of birds.
The inset highlights the cases corresponding to the configurations shown in Fig.~\ref{fig:3}(a).
\label{fig:3}
}
\end{figure}

Figure~\ref{fig:3}(a) shows the steady-state configurations obtained numerically 
for homogeneous formations with different numbers of birds. 
All configurations exhibit a U-shaped structure 
with hierarchical streamwise spacing along each wing. 
Specifically, the spacing between adjacent birds decreases 
monotonically toward the leader, satisfying the ordering relation
\begin{equation}
x_2 - x_1 > x_3 - x_2 > \cdots > x_n - x_{n-1}.
\label{eq:ordering_numerical_confirmation}
\end{equation}
The numerically observed spacing is consistent with the analytical prediction 
in Eq.~(\ref{eq:linear_general_x}). 
Moreover, the results obtained from the all-to-all interaction (solid lines)
and the nearest-neighbor approximation (dashed lines) are nearly indistinguishable, 
indicating that the formation geometry is predominantly governed 
by nearest-neighbor interactions.

As the total number of birds $N$ increases, 
the steady formation becomes progressively wider, 
with both the lateral span and the streamwise extent 
increasing systematically. 
Meanwhile, the hierarchical spacing structure remains preserved, 
so that larger formations correspond to geometrically expanded 
yet structurally similar steady configurations.

Figure~\ref{fig:3}(b) shows the dependence of the steady-state velocity 
$U_{\mathrm{steady}}$ on the number of birds $N$. 
As $N$ increases, $U_{\mathrm{steady}}$ increases monotonically, 
demonstrating that formation flight enhances the overall flight speed 
through aerodynamic interactions among the birds. 
However, the incremental increase becomes progressively smaller 
for larger $N$, indicating a diminishing marginal benefit 
as the flock size grows.

Figure~\ref{fig:3}(c) presents the average drag reduction rate \(\eta\) 
as a function of \(N\). 
In the nearest-neighbor approximation, the drag reduction rate of bird \(j\) is defined as
\begin{equation}
\eta_j
=
-\frac{1}{D_0}
\sum_{i\in\mathcal{N}_j} D_{ij},
\end{equation}
where \(\mathcal{N}_j\) denotes the nearest neighbors
of bird \(j\), and \(D_0\) is the drag of a solitary bird at its cruising speed.
The quantity shown in Fig.~\ref{fig:3}(c) is the average drag reduction rate in the steady state formation,
\begin{equation}
\eta = \frac{1}{N}\sum_{j=1}^{N} \eta_j.
\end{equation}
The monotonic increase of $\eta$ with $N$ provides a direct 
mechanical explanation for the velocity enhancement observed 
in Fig.~\ref{fig:3}(b): reduced aerodynamic drag allows the formation 
to sustain a higher steady flight speed under the same thrust condition. 
Although $\eta$ continues to increase with $N$, 
its growth rate decreases for large formations, 
reflecting the finite effective range of aerodynamic interactions.

The trends observed in Fig.~\ref{fig:3} can be related to previous
theoretical studies of formation flight.
In particular, Sugimoto~\cite{sugimoto2001theoretical}
analyzed formation flight as a nonlinear self-organizing phenomenon
using an elliptically loaded lifting-line model,
and found stable steady formations with a U-shaped geometry.
The present study shares the same dynamical viewpoint that formation
geometry emerges as a steady state of aerodynamic interactions, but
adopts a more reduced description based on an equivalent
horseshoe-vortex representation and simplified local longitudinal
dynamics.
Within this framework, the homogeneous steady formations obtained here
exhibit a U-shaped structure with hierarchical streamwise spacing
[Eq.~(\ref{eq:ordering_numerical_confirmation})],
consistent with the U-shaped geometry reported by Sugimoto.
In addition, while Sugimoto focused on the existence, stability, and
self-organization of homogeneous formations, the present work extends
the dynamical analysis to localized inhomogeneity and reveals a strong
position dependence of the stability range.

\section{\label{sec:inhomogeneous} Numerical Results for Inhomogeneous Formations}

In the previous sections, we analyzed homogeneous formations
for a general odd number of birds $N=2n-1$.
For the numerical investigation of inhomogeneous formations,
we focus on a representative flock size of $N=13$,
which is close to the flock sizes reported in previous studies
of V-shaped formations~\cite{portugal2014upwash,mirzaeinia2020analytical}.

Inhomogeneity is introduced by modifying the span ratio $\lambda$
of a single bird, so that its semi-wingspan is $b_i=\lambda b$,
while keeping all other birds identical.
Owing to the left--right symmetry of the formation, we choose the single
modified bird from the left wing, with its index taken as $i=1,\dots,7$,
where $i=7$ corresponds to the leader.
For an adjacent pair with semi-wingspans $b_i$ and $b_{i+1}$, 
we set the transverse center-to-center separation as
\[
y_{i+1,i} = b_i+b_{i+1}+s_{\rm opt},
\]
where $s_{\rm opt}=(\pi/4-1)b$.
Thus, the wingtip overlap is kept fixed at $|s_{\rm opt}|$ when the wingspan of one bird is varied.
For the inhomogeneous simulations,
the initial condition is constructed from the homogeneous reference configuration,
with the longitudinal positions retained from the homogeneous case
and the transverse separations modified according to the inhomogeneous wingspans.
All birds are assigned the same initial velocity, $U_i(0)=U_0$.

\begin{figure}[t]
\includegraphics[width=0.8\columnwidth]{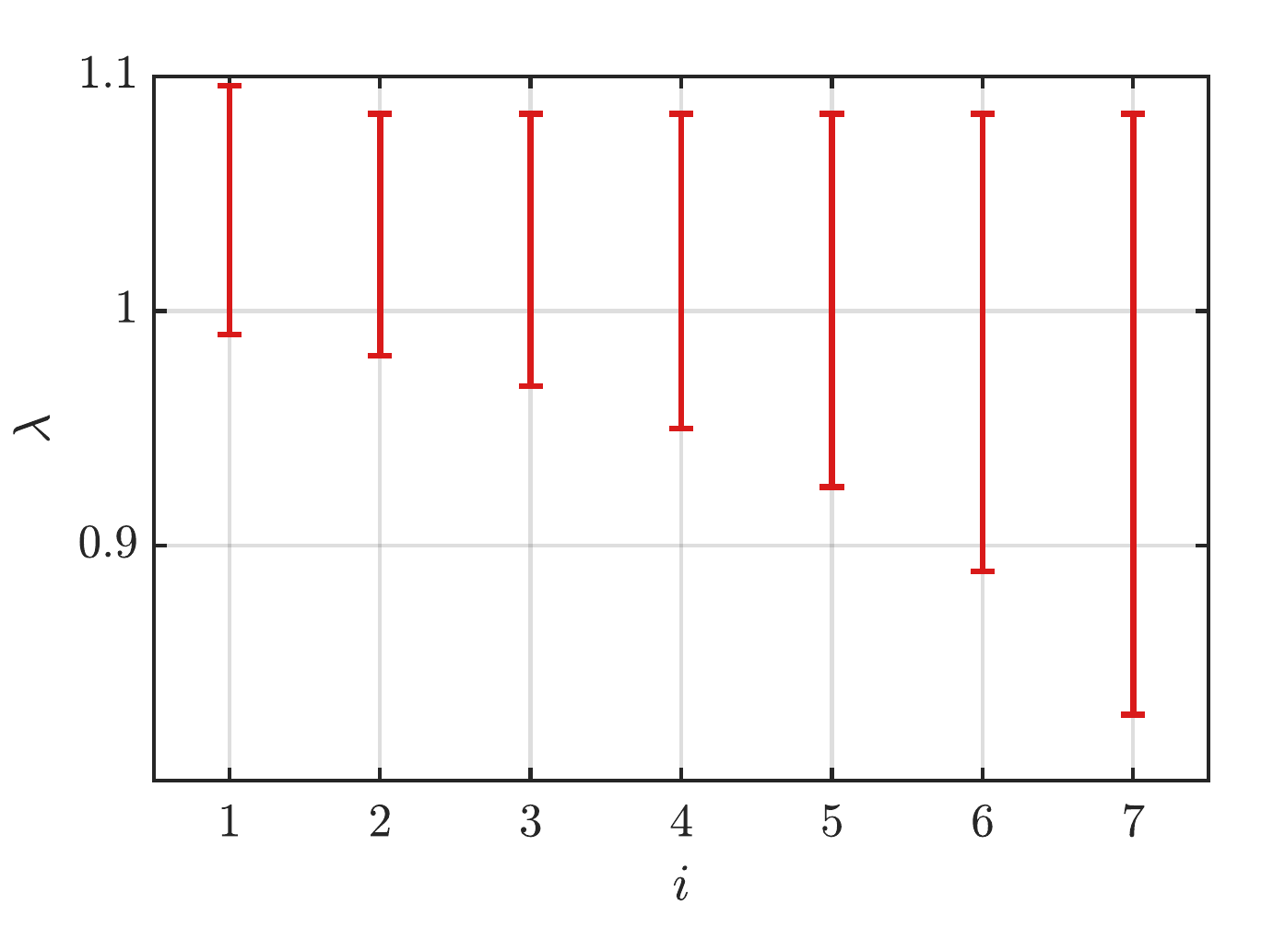}
\caption{
Stability diagram in the $(i,\lambda)$ plane for a formation of 13 birds.
The red vertical lines indicate the interval of \(\lambda\)
for which the formation remains stable.
\label{fig:4}
}
\end{figure}

Fig.~\ref{fig:4} shows the stable interval of $\lambda$
as a function of the index $i$ of the modified bird.
The range $0.5\le\lambda\le1.5$
is chosen following the parameter range considered in Ref.~\cite{cattivelli2011modeling}, 
where the wingspan-related aerodynamic properties of each bird
were randomly varied between $0.5$ and $1.5$
to investigate heterogeneous formations. Their simulations showed that 
V formations can still emerge under such substantial individual variability. 
In contrast, the present study introduces inhomogeneity by modifying the 
span ratio of a single bird while keeping the others identical, allowing 
us to systematically examine the dynamical stability of the formation 
under a localized inhomogeneity.

Because the aerodynamic and dynamical properties scale nonlinearly with $b$ 
($m\propto b^3$, $T\propto b^4$), this interval represents a substantial 
variation in thrust and drag characteristics. It therefore allows us to 
investigate the stability of the formation over a broad range of 
aerodynamic contrasts.

In the numerical simulations, stability is quantified by the standard 
deviation of the bird velocities,
\begin{equation}
\sigma_U
=
\sqrt{\langle U^2 \rangle - \langle U \rangle^2},
\end{equation}
where
\begin{equation}
\langle U \rangle = \frac{1}{N}\sum_{j=1}^{N} U_j,
\qquad
\langle U^2 \rangle = \frac{1}{N}\sum_{j=1}^{N} U_j^2.
\end{equation}

The quantity $\sigma_U$ is evaluated at the final simulation time 
$t=t_{\max}=10^5$, by which time transient dynamics have fully decayed.
A formation is considered stable if
$\sigma_U<\varepsilon$,
with $\varepsilon=10^{-4}$,
and unstable otherwise.
The threshold value was chosen sufficiently smaller
than the typical steady-state velocity scale.
The red vertical lines therefore indicate
the parameter interval
in which stable collective motion is maintained.

The results clearly demonstrate that the stability range of $\lambda$ 
depends strongly on the spatial position of the inhomogeneous bird. 
The admissible interval of \(\lambda\) becomes significantly broader
when the modified bird is located closer to the leader (\(i\to 7\)).

This trend indicates that the central region of the formation is 
dynamically more robust against inhomogeneity. A bird located near 
the leader experiences stronger aerodynamic coupling from neighboring 
individuals, which enhances the collective stability of the configuration. 
By contrast, perturbations introduced near the boundary are less 
effectively constrained, making the formation more susceptible 
to instability.

\begin{figure*}[t]
  \centering
  \includegraphics[width=\textwidth]{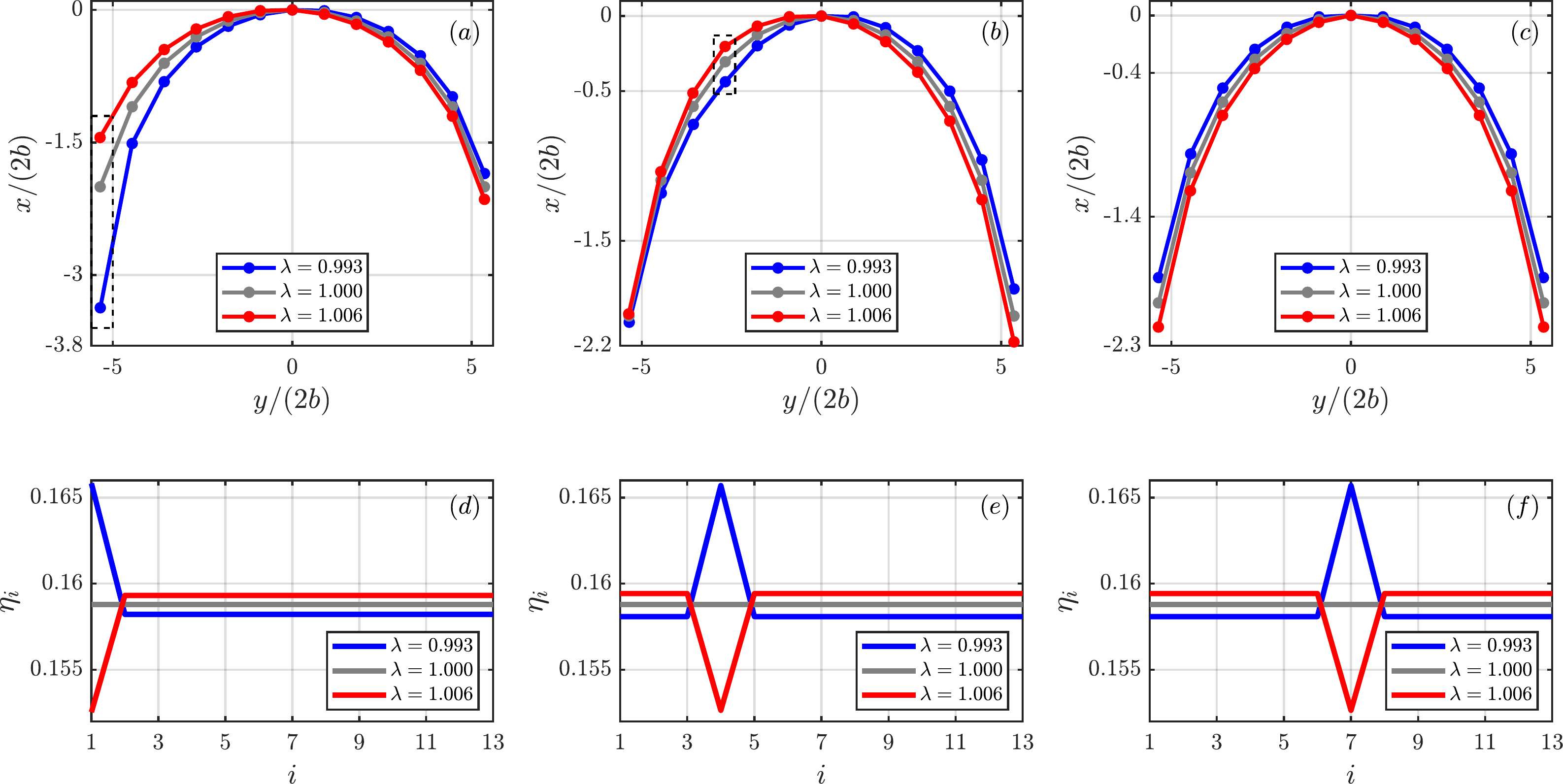}
  \caption{
    Steady configurations and drag reduction rates
    in inhomogeneous formations with $N=13$.
    (a)--(c) Steady-state configurations. The coordinates are scaled by the reference wingspan $2b$ and shifted so that the leader is located at the origin.
    The modified bird is located at
    (a) the outermost bird ($i=1$),
    (b) an intermediate bird on the left wing ($i=4$), and
    (c) the leader ($i=7$).
    In (a) and (b), the modified bird is highlighted by a black dashed box.
    Different colors correspond to $\lambda<1$, $\lambda=1$, and $\lambda>1$.
    (d)--(f) Corresponding individual drag reduction rates $\eta_i$.
    }
  \label{fig:5}
\end{figure*}

Fig.~\ref{fig:5}(a)--(c) show representative steady-state configurations 
for three representative positions
of the inhomogeneous bird:
the outermost bird ($i=1$), 
an intermediate bird on the left wing ($i=4$), 
and the leader ($i=7$), respectively. 
For each position,
representative cases with
$\lambda<1$, $\lambda=1$, and $\lambda>1$
are shown.
When the inhomogeneity is introduced at the outermost bird
or at an intermediate position [Figs.~\ref{fig:5}(a) and \ref{fig:5}(b)], 
a clear positional shift is observed. 
For $\lambda>1$, the modified bird moves forward 
relative to the homogeneous configuration ($\lambda=1$), 
whereas for $\lambda<1$, it lags behind. 
The deformation remains primarily localized
near the inhomogeneous bird,
while the global formation geometry
remains largely unchanged.
In contrast, when the leader is modified [Fig.~\ref{fig:5}(c)], 
the geometric response of the formation becomes global. 
As $\lambda$ increases, the formation becomes sharper, 
with a narrower opening angle. 
Conversely, for $\lambda<1$, 
the formation becomes flatter
with a broader opening angle.
Thus, altering the leader modifies the global geometry 
of the formation rather than inducing a localized displacement.

Fig.~\ref{fig:5}(d)--(f) show
the individual drag reduction rates $\eta_i$
corresponding to
Fig.~\ref{fig:5}(a)--(c), respectively.
In all cases, the most pronounced change in $\eta_i$
occurs at the modified bird itself. 
For $\lambda>1$, the modified bird experiences 
a higher drag reduction than in the homogeneous case, 
whereas for $\lambda<1$, its drag reduction decreases. 

\section{\label{sec:discussion}Discussion and Conclusion}

The position dependence of stability can be understood
from the local structure of aerodynamic interactions.
In the present model, these interactions are short-ranged,
so that the effective dynamical coupling varies across the formation.

The steady formations obtained here also differ
from those discussed in previous studies.
In the homogeneous case, the present model yields
a U-shaped formation with hierarchical streamwise spacing.
While U-shaped configurations have also been reported in earlier work
\cite{sugimoto2001theoretical,kawabe2007study},
the present results arise from a dynamical self-organization process
driven by local aerodynamic interactions.
This result suggests that
the resulting formation geometry
depends not only on aerodynamic or energetic considerations,
but also on the dynamical processes
that determine the steady-state configuration
and its stability.

The present results also extend previous studies
of heterogeneous formations~\cite{hummel1983aerodynamic,cattivelli2011modeling}.
While earlier works showed that coherent formation flight can persist
despite individual variability,
our analysis demonstrates that the position
of a localized inhomogeneity is a key factor
controlling both the resulting geometry and its stability.
This highlights the importance of considering not only
the magnitude of heterogeneity,
but also its spatial distribution within the formation.

In summary, we investigated the effects of inhomogeneity
on the formation flight of migrating birds
using a lifting-line model with an equivalent horseshoe-vortex
representation coupled to longitudinal flight dynamics.
For homogeneous formations,
we obtained steady states with hierarchical streamwise spacing,
in which the streamwise distance between adjacent birds decreases
toward the leader, resulting in a U-shaped formation.
For inhomogeneous formations,
we showed that the stability range of the span ratio
depends strongly on the position of the modified bird:
perturbations near the leader are accommodated more robustly,
whereas those near the outermost bird are less stable.
Moreover, inhomogeneity away from the leader mainly induces
localized deformation,
whereas modifying the leader causes a global
reorganization of the formation geometry.

These findings highlight the importance of a dynamical perspective
for understanding how collective flight structures emerge,
are maintained, and respond to localized individual differences.

\section*{Data Availability}

The data and codes that support the findings of this study are openly
available in Zenodo~\cite{zenodo_data}.

\begin{acknowledgments}
This work was supported by the JSPS KAKENHI {24K06895} for NU.\par
\end{acknowledgments}


\bibliography{jiang2026localized}%

@PREAMBLE{
 "\providecommand{\noopsort}[1]{}" 
 # "\providecommand{\singleletter}[1]{#1}%" 
}

@article{voelkl2017relation,
  author = {Voelkl, B. and Fritz, J.},
  title = {Relation between travel strategy and social organization of migrating birds with special consideration of formation flight in the northern bald ibis},
  journal = {Philos. Trans. R. Soc. B},
  volume = {372},
  number = {1727},
  pages = {20160235},
  year = {2017},
  doi = {10.1098/rstb.2016.0235}
}

@article{andersson2004kin,
  author  = {Andersson, Malte and Wallander, Johan},
  title   = {Kin selection and reciprocity in flight formation?},
  journal = {Behavioral Ecology},
  volume  = {15},
  number  = {1},
  pages   = {158--162},
  year    = {2004},
  doi     = {10.1093/beheco/arg109}
}

@article{bajec2009organized,
  author = {Bajec, I. L. and Heppner, F. H.},
  title = {Organized flight in birds},
  journal = {Anim. Behav.},
  volume = {78},
  number = {4},
  pages = {777--789},
  year = {2009},
  doi = {10.1016/j.anbehav.2009.07.007}
}

@article{ward1978formation,
  author = {Ward, P.},
  title = {Formation-flying of birds as advertisement behaviour},
  journal = {Anim. Behav.},
  volume = {26},
  pages = {1273},
  year = {1978},
  doi = {10.1016/0003-3472(78)90118-5}
}

@article{heppner1974avian,
  author = {Heppner, F. H.},
  title = {Avian Flight Formations},
  journal = {Bird-Banding},
  volume = {45},
  number = {2},
  pages = {160--169},
  year = {1974},
  doi = {10.2307/4512025}
}

@article{nathan2008vlike,
  author = {Nathan, A. and Barbosa, V. C.},
  title = {V-like formations in flocks of artificial birds},
  journal = {Artif. Life},
  volume = {14},
  number = {2},
  pages = {179--188},
  year = {2008},
  doi = {10.1162/artl.2008.14.2.179}
}

@article{sewatkar2010first,
  author = {Sewatkar, C. M. and Sharma, A. and Agrawal, A.},
  title = {A first attempt to numerically compute forces on birds in {V} formation},
  journal = {Artif. Life},
  volume = {16},
  number = {3},
  pages = {245--258},
  year = {2010},
  doi = {10.1162/artl_a_00005}
}

@article{beauchamp2011long,
  author = {Beauchamp, G.},
  title = {Long-distance migrating species of birds travel in larger groups},
  journal = {Biol. Lett.},
  volume = {7},
  number = {5},
  pages = {692--694},
  year = {2011},
  doi = {10.1098/rsbl.2011.0243}
}

@article{lissaman1970formation,
  author = {P. B. S. Lissaman and C. A. Shollenberger},
  title = {Formation Flight of Birds},
  journal = {Science},
  volume = {168},
  pages = {1003--1005},
  year = {1970},
  doi = {10.1126/science.168.3934.1003}
}

@article{badgerow1981energy,
  author = {Badgerow, J. P. and Hainsworth, F. R.},
  title = {Energy savings through formation flight? A re-examination of the vee formation},
  journal = {J. Theor. Biol.},
  volume = {93},
  pages = {41--52},
  year = {1981}
}

@article{weimerskirch2001energy,
  author  = {Weimerskirch, Henri and Martin, Julien and Clerquin, Yannick and Alexandre, Peggy and Jiraskova, Sarka},
  title   = {Energy saving in flight formation},
  journal = {Nature},
  volume  = {413},
  pages   = {697--698},
  year    = {2001},
  doi     = {10.1038/35099670}
}

@article{heppner1985visual,
  author  = {Frank H. Heppner and Jeffrey L. Convissar and Dennis E. Moonan and John G. T. Anderson},
  title   = {Visual Angle and Formation Flight in Canada Geese (Branta canadensis)},
  journal = {The Auk},
  volume  = {102},
  number  = {1},
  pages   = {195--198},
  year    = {1985},
  doi     = {10.2307/4086847}
}

@inproceedings{li2017v-shaped,
  author    = {X. Li and Y. Tan and J. Fu and I. Mareels},
  title     = {On V-Shaped Flight Formation of Bird Flocks with Visual Communication Constraints},
  booktitle = {Proceedings of the 2017 13th IEEE International Conference on Control \& Automation (ICCA)},
  year      = {2017},
  pages     = {513--518},
  address   = {Ohrid, Macedonia},
  month     = jul,
  publisher = {IEEE},
  doi       = {https://doi.org/10.1109/ICCA.2017.8003113}
}

@inproceedings{seiler2002analysis,
  author = {Seiler, Peter and Pant, Amit and Hedrick, J. Karl},
  title = {Analysis of bird formations},
  booktitle = {Proc. 41st IEEE Conf. Decision and Control},
  pages = {118--123},
  year = {2002},
  address = {Las Vegas, NV, USA},
  doi = {10.1109/CDC.2002.1184478}
}

@article{vine1971risk,
  author  = {I. Vine},
  title   = {Risk of visual detection and pursuit by predator and the selective advantage of flocking behaviour},
  journal = {Journal of Theoretical Biology},
  volume  = {30},
  number  = {2},
  pages   = {405--422},
  year    = {1971}
}

@article{cattivelli2011modeling,
  author  = {Cattivelli, Fabio S. and Sayed, Ali H.},
  title   = {Modeling bird flight formations using diffusion adaptation},
  journal = {IEEE Transactions on Signal Processing},
  volume  = {59},
  number  = {5},
  pages   = {2038--2051},
  year    = {2011},
  doi     = {10.1109/TSP.2011.2105480}
}

@article{cutts1994energy,
  author  = {C. J. Cutts and J. R. Speakman},
  title   = {Energy savings in formation flight of pink-footed geese},
  journal = {Journal of Experimental Biology},
  volume  = {189},
  pages   = {251--261},
  year    = {1994}
}

@article{portugal2014upwash,
  author  = {Portugal, Steven J. and Hubel, Tatjana Y. and Fritz, Johannes and Heese, Stefanie and Trobe, Daniela and Voelkl, Bernhard and Hailes, Stephen and Wilson, Alan M. and Usherwood, James R.},
  title   = {Upwash exploitation and downwash avoidance by flap phasing in ibis formation flight},
  journal = {Nature},
  volume  = {505},
  number  = {7483},
  pages   = {399--402},
  year    = {2014},
  doi     = {10.1038/nature12939}
}

@book{dorst1962migration,
  author = {Dorst, J.},
  title = {The Migration of Birds},
  publisher = {Houghton Mifflin Company},
  address = {Boston, MA},
  year = {1962}
}

@incollection{hamilton1967social,
  author    = {W. J. Hamilton III},
  title     = {Social aspects of bird orientation mechanisms},
  booktitle = {Animal Orientation and Migration},
  editor    = {R. M. Storm},
  publisher = {Oregon State University Press},
  address   = {Corvallis, OR},
  year      = {1967},
  pages     = {57--71}
}

@article{hummel1995formation,
  author  = {D. Hummel},
  title   = {Formation flight as an energy-saving mechanism},
  journal = {Israel Journal of Zoology},
  volume  = {41},
  number  = {3},
  pages   = {261--278},
  year    = {1995}
}

@article{corcoran2019compound,
  author = {Corcoran, Aaron J. and Hedrick, Tyson L.},
  title = {Compound-V formations in shorebird flocks},
  journal = {eLife},
  volume = {8},
  pages = {e45071},
  year = {2019},
  doi = {10.7554/eLife.45071}
}

@article{voelkl2015matching,
  author = {Voelkl, Bernhard and Portugal, Steven J. and Uns{\"o}ld, Markus and Usherwood, James R. and Wilson, Alan M. and Fritz, Johannes},
  title = {Matching times of leading and following suggest cooperation through direct reciprocity during V-formation flight in ibis},
  journal = {Proc. Natl. Acad. Sci. U.S.A.},
  volume = {112},
  number = {7},
  pages = {2115--2120},
  year = {2015},
  doi = {10.1073/pnas.1413589112}
}

@article{perinot2023characterization,
  author = {Perinot, Elisa and Fritz, Johannes and Fusani, Leonida and Voelkl, Bernhard and Nobile, Marco S.},
  title = {Characterization of bird formations using fuzzy modelling},
  journal = {J. R. Soc. Interface},
  volume = {20},
  pages = {20220798},
  year = {2023},
  doi = {10.1098/rsif.2022.0798}
}

@article{billingsley2021role,
  author = {Billingsley, E. and Ghommem, M. and Vasconcellos, R. and Abdelkefi, A.},
  title = {Role of Active Morphing in the Aerodynamic Performance of Flapping Wings in Formation Flight},
  journal = {Drones},
  volume = {5},
  number = {3},
  pages = {90},
  year = {2021},
  doi = {10.3390/drones5030090}
}

@article{maeng2013modeling,
  author  = {Maeng, Joo-Sung and Park, Jae-Hyung and Jang, Seong-Min and Han, Seog-Young},
  title   = {A modeling approach to energy savings of flying Canada geese using computational fluid dynamics},
  journal = {J. Theor. Biol.},
  volume  = {320},
  pages   = {76--85},
  year    = {2013},
  doi     = {10.1016/j.jtbi.2012.11.032}
}

@article{beaumont2023aerodynamic,
  author  = {Beaumont, F. and Murer, S. and Bogard, F. and Polidori, G.},
  title   = {Aerodynamic Interaction of Migratory Birds in Gliding Flight},
  journal = {Fluids},
  volume  = {8},
  pages   = {50},
  year    = {2023},
  doi     = {10.3390/fluids8020050}
}

@article{beaumont2025aerodymic,
  author  = {Fabien Beaumont and S{\'e}bastien Murer and Fabien Bogard and Guillaume Polidori},
  title   = {Aerodynamic mechanisms behind energy efficiency in migratory bird formations},
  journal = {Physics of Fluids},
  volume  = {37},
  pages   = {025202},
  year    = {2025},
  doi     = {10.1063/5.0252553}
}

@article{beaumont2025aerodynamics,
  author  = {Beaumont, Fabien and Murer, S{\'e}bastien and Bogard, Fabien and Polidori, Guillaume},
  title   = {Aerodynamics of Flight Formations in Birds: A Quest for Energy Efficiency},
  journal = {Birds},
  volume  = {6},
  number  = {2},
  pages   = {15},
  year    = {2025},
  doi     = {10.3390/birds6020015}
}

@article{vicsek1995novel,
  author = {Vicsek, T. and Czir{\'o}k, A. and Ben-Jacob, E. and Cohen, I. and Shochet, O.},
  title = {Novel Type of Phase Transition in a System of Self-Driven Particles},
  journal = {Phys. Rev. Lett.},
  volume = {75},
  pages = {1226--1229},
  year = {1995},
  doi = {10.1103/PhysRevLett.75.1226}
}

@article{couzin2002collective,
  author = {Couzin, I. D. and Krause, J. and James, R. and Ruxton, G. D. and Franks, N. R.},
  title = {Collective Memory and Spatial Sorting in Animal Groups},
  journal = {J. Theor. Biol.},
  volume = {218},
  number = {1},
  pages = {1--11},
  year = {2002},
  doi = {10.1006/jtbi.2002.3065}
}

@inproceedings{reynolds1987,
  author = {Reynolds, C. W.},
  title = {Flocks, Herds and Schools: A Distributed Behavioral Model},
  booktitle = {Proc. 14th Annual Conference on Computer Graphics and Interactive Techniques (SIGGRAPH)},
  pages = {25--34},
  year = {1987},
  doi = {10.1145/37401.37406}
}

@article{cucker2007emergent,
  author = {Cucker, F. and Smale, S.},
  title = {Emergent Behavior in Flocks},
  journal = {IEEE Trans. Autom. Control},
  volume = {52},
  number = {5},
  pages = {852--862},
  year = {2007},
  doi = {10.1109/TAC.2007.895842}
}

@article{herbertRead2016understanding,
  author = {Herbert-Read, J. E.},
  title = {Understanding How Animal Groups Achieve Coordinated Movement},
  journal = {J. Exp. Biol.},
  volume = {219},
  number = {19},
  pages = {2971--2983},
  year = {2016},
  doi = {10.1242/jeb.129411}
}

@inproceedings{cattivelli2009self,
  author = {Cattivelli, F. and Sayed, A. H.},
  title = {Self-organization in bird flight formations using diffusion adaptation},
  booktitle = {Proc. 3rd IEEE Int. Workshop on Computational Advances in Multi-Sensor Adaptive Processing (CAMSAP)},
  year = {2009},
  pages = {257--260},
  doi = {10.1109/CAMSAP.2009.5413237}
}

@article{sugimoto2001theoretical,
  author  = {Sugimoto, Norio},
  title   = {Theoretical study of formation flight of birds},
  journal = {Journal of Theoretical Biology},
  volume  = {211},
  number  = {2},
  pages   = {181--192},
  year    = {2001},
  doi     = {10.1006/jtbi.2001.2333}
}

@inproceedings{thien2007effects,
  author = {Thien, H. P. and Moelyadi, M. A. and Muhammad, H.},
  title = {Effects of Leader's Position and Shape on Aerodynamic Performances of {V} Flight Formation},
  booktitle = {Proc. International Conference on Intelligent Unmanned Systems (ICIUS)},
  year = {2007},
  address = {Bali, Indonesia},
  pages = {A008}
}

@article{hummel1983aerodynamic,
  author  = {Hummel, D.},
  title   = {Aerodynamic aspects of formation flight in birds},
  journal = {Journal of Theoretical Biology},
  volume  = {104},
  number  = {3},
  pages   = {321--347},
  year    = {1983},
  publisher = {Academic Press}
}

@inproceedings{mirzaeinia2020flocking,
  author    = {Mirzaeinia, Amir and Mirzaeinia, Mehdi and Hassanalian, Mostafa},
  title     = {Flocking of V-shaped and Echelon Northern Bald Ibises with Different Wingspans: Repositioning and Energy Saving},
  booktitle = {AIAA SciTech 2020 Forum},
  year      = {2020},
  pages     = {2020-0052},
  doi       = {10.2514/6.2020-0052}
}

@article{kawabe2007study,
  author  = {Hiroyasu Kawabe},
  title   = {A Study on Optimal Pattern and Leader Shift of Formation Flight},
  journal = {Transactions of the Japan Society for Aeronautical and Space Sciences},
  volume  = {50},
  number  = {168},
  pages   = {134--140},
  year    = {2007},
  doi     = {10.2322/tjsass.50.134}
}

@book{pennycuick2008modelling,
  author    = {Colin J. Pennycuick},
  title     = {Modelling the Flying Bird},
  year      = {2008},
  publisher = {Academic Press},
  address   = {London},
  series    = {Theoretical Ecology Series}
}

@article{shyy2010recent,
  title={Recent progress in flapping wing aerodynamics and aeroelasticity},
  author={Wei Shyy and Hikaru Aono and Satish Kumar Chimakurthi and Pat Trizila and Chang-kwon Kang and Carlos E. S. Cesnik and Hao Liu},
  journal={Progress in Aerospace Sciences},
  year={2010},
  volume={46},
  pages={284-327},
  url={https://api.semanticscholar.org/CorpusID:110014894}
}

@article{mirzaeinia2020analytical,
  author  = {Mirzaeinia, A. and Verma, S. and Shukla, S.},
  title   = {Analytical modeling of aerodynamic interaction in V-formation flight of birds},
  journal = {Acta Mechanica Sinica},
  volume  = {36},
  pages   = {841--852},
  year    = {2020},
  doi     = {10.1007/s10409-020-00978-6}
}

@article{Tobalske2007,
  author  = {Bret W. Tobalske and Kenneth P. Dial},
  title   = {Aerodynamics of wing-assisted incline running in birds},
  journal = {Journal of Experimental Biology},
  year    = {2007},
  volume  = {210},
  pages   = {1742--1751},
  doi     = {10.1242/jeb.001701}
}

@article{Hainsworth1987,
  author = {Hainsworth, F. R.},
  title = {Precision and dynamics of positioning by Canada geese flying in formation},
  journal = {J. Exp. Biol.},
  volume = {128},
  number = {1},
  pages = {445--462},
  year = {1987}
}

@misc{zenodo_data,
  author       = {Jiang, Hui and Uchida, Nariya},
  title        ={Dataset and code for: Localized inhomogeneity and position-dependent stability of migratory bird formations, https://doi.org/10.5281/zenodo.20070621},
  month        = apr,
  publisher = {Zenodo},
  doi          = {10.5281/zenodo.20070621},
  url          = {https://doi.org/10.5281/zenodo.20070621}
}
\end{document}